\documentclass[12pt,twoside]{article} 
\usepackage{fleqn,espcrc1} 
\usepackage{epsf,subfigure} 
\usepackage{graphicx} 
\usepackage[figuresright]{rotating} 
  
\newcommand{\AmS}{{\protect\the\textfont2  
  A\kern-.1667em\lower.5ex\hbox{M}\kern-.125emS}}  
  
\title{Relativistic wave functions and energies for nonzero 
angular momentum states in light-front dynamics} \author{V.A. Karmanov\address{Lebedev Physical Institute, Leninsky Prospekt 53, 117924 Moscow, Russia}, 
J. Carbonell$^{\rm b}$ 
and 
M. Mangin-Brinet\address{Institut des Sciences   Nucl\'eaires, 53 av. des Martyrs, 38026 Grenoble Cedex, France}}   
         
\begin{document}  
   
\maketitle   
Light-front dynamics (LFD) is a powerful approach to the theory of   
relativistic composite systems (hadrons in the quark models and   
relativistic nucleons in nuclei). Its explicitly covariant version   
\cite{cdkm} has been recently applied with success  
\cite{ck99} to describe the new CEBAF/TJNAF data on the deuteron   
electromagnetic form factors. The solutions used in \cite{ck99} 
were however not obtained from solving exactly the LFD equations but by means 
of a perturbative calculation with respect to the non relativistic wave 
function.  
Since, a consequent effort has been made to obtain exact solutions of LFD  
equations. The first results concerning $J=0$ states in a scalar model  
have been published in \cite{MC_00}.   
The construction of $J \ne 0$ states in LFD is complicated by the two 
following facts.  
First, the generators of the spatial rotations contain interaction 
and are thus difficult to handle.  
Second, one is always forced to work in a truncated  
Fock space, and consequently, the Poincar\'e group commutation  
relations between the generators -- ensuring the correct properties 
of the state vector under rotation -- are in practice destroyed. 
In the standard approach, with the light-front plane defined  
as $t+z=0$, this violation of rotational invariance  
manifests by the fact that the energy depends  
on the angular momentum projection on $z$-axis \cite{cmp}. 
 
We present here a method to construct $J\ne0$ states in the explicitly 
covariant formulation of LFD and show how it leads to a restoration of  
rotational invariance.    
In this approach \cite{cdkm} the equation for    
relativistic two-body bound state wave function   
$\psi$ reads:  
\begin{equation}\label{eq1}  
[4(\vec{k}\,^2 +m^2)-M^2] \psi(\vec{k},\hat{n})=-{m^2 \over 2   
\pi^3} \int {d^3k' \over \varepsilon_{k'}   
}V(\vec{k},\vec{k'},\hat{n},M^2)\psi(\vec{k'},\hat{n}),   
\end{equation}   
where $V$ is the interaction kernel,    
$\hat{n}$ the  three-dimensional unit vector determining   
the orientation of the light-front plane, $M$ the total mass of the system and 
$\varepsilon_{k}=\sqrt{m^2+{\vec{k}}^2}$.  In general, the wave function  
depends both on the relative momentum $\vec{k}$ and $\hat{n}$.   
In order to solve the problem, we replace the dynamical angular momentum 
operator by the kinematical one:    
\begin{equation}\label{eq2}  
\vec{J} =                                                               
-i[\vec{k}\times \partial/\partial\vec{k}\,] -i[\hat{n}\times   
\partial/\partial\hat{n}].  
\end{equation}  
The eigenstates of $\vec{J}^2$ are constructed from vectors   
$\vec{k},\hat{n}$ -- taken on equal ground -- by using the standard 
techniques of angular momentum theory. 
It turns out that the kernel and $\vec{J}$ commute not only with each other,  
but also   
with the operator $A^2=(\hat{n}\cdot\vec{J})^2$. Hence, the   
solutions of (\ref{eq1}) are labelled by their mass, angular momentum and  
projection on $z$-axis, but also by the eigenvalues   
$a^2=0,1,\ldots,J^2$ of $A^2$. In a truncated Fock space, the states with 
different $a$ correspond to different masses $M_a$.    
In the  $J^{\pi}=1^-$ case for instance, the solutions  
$\vec{\psi}_{a}$ with $a=0,1$ can be written in the form:  
$$ 
\vec{\psi}_{a}(\vec{k},\hat{n})=  
\chi_{a}(\hat{k},\hat{n}) g_{a}(k,z),  
$$  
where $\chi_{a}$ are the normalized   
eigenfunctions of the operator $A^2$:    
$$A^2\chi_a(\hat{k},\hat{n}) =a^2   
\chi_a(\hat{k},\hat{n}),  
\qquad  
\chi_0(\hat{k},\hat{n})=3z\hat{n},  
\qquad   
\chi_1(\hat{k},\hat{n})  
=\frac{3\sqrt{2}}{2}(\hat{k}-z\hat{n}),  
$$  
$z=\hat{k}\cdot\hat{n}$ and functions $g_{a}$ satisfy the equations 
\begin{eqnarray*}  
&&[4(\vec{k}\,^2 +m^2)-M_0^2] zg_0(k,z)=-\frac{m^2}{2 \pi^3}   
\int \frac{d^3 k'}{\varepsilon_{k'}}  
V(\vec{k},\vec{k}\,',\hat{n},M_0^2)z'g_0(k',z'),  
\nonumber\\  
&&[4(\vec{k}\,^2 +m^2)-M_1^2](1-z^2)g_1(k,z) =-\frac{m^2}{2 \pi^3}   
\int \frac{d^3k'}{ \varepsilon_{k'}}  
V(\vec{k},\vec{k}\,',\hat{n},M_1^2)  
(\hat{k}\cdot\hat{k}'-zz') g_1(k',z').  
\end{eqnarray*}  
obtained by inserting $\vec{\psi}_{0,1}$ into (\ref{eq1}).  
As mentioned above, due to the Fock space truncation, the masses $M_{0,1}$  
determined in this way differ from each other.  
The key point of our method is based on the fact that, in order to  
ensure the equivalence between the dynamical angular momentum operator   
and the kinematical one (\ref{eq2}), the physical state has  
to satisfy the so called angular condition \cite{VAK_82}.
This condition imposes the physical wave function to be a superposition 
of eigenstates of $A^2$:   
\begin{equation}\label{eq2a}  
\psi(\vec{k},\hat{n})=\sum_a c_a \psi_a(\vec{k},\hat{n}) =  
 c_0\chi_0(\hat{k},\hat{n})g_0(k,z)+  
 c_1\chi_1(\hat{k},\hat{n})g_1(k,z)  
\end{equation}  
The unknown coefficients are in principle determined by the angular condition
itself but they can be unambiguously fixed by imposing that in the limit
$k\to 0$, the solution does not depend on $\hat{n}$.
The corresponding mass is then given by    
\begin{equation}\label{eq3} 
M^2=\sum_a c^2_a M^2_a.   
\end{equation}   
The validity of this solution has been checked in the Wick-Cutkosky   
model, i.e.  for two scalar particles interacting by a massless scalar 
exchange. For $J=1^-$ state, the values of $M_{a}$, $M$ given by (\ref{eq3}) 
and the mass of the 2s states are shown in   
Figure (a) as functions of the coupling constant $\alpha$.  One can see 
that despite the big split between $M_0$ and $M_1$, $M$ for the 1p state 
is very close to the  mass of the 2s one, which in its turn is very close 
to the results obtained by solving Bethe-Salpeter equation (denoted by dots).
The curves for $M$ (solid line) and the 2s mass (long-dashed line) are 
hardly distinguishable. This means that our solution   
restores quite accurately the 2s-1p degeneracy of the relativistic Coulomb   
problem which in the Bethe-Salpeter approach is an exact result. The   
same situation also takes place in Figure (b) for 3s-2p states.  
It is worth noticing that in the whole range of coupling constants considered
-- implying large binding energies -- the coefficients $c_{0,1}$ are 
very close to $\sqrt{1/3}$ and $\sqrt{2/3}$.
To summarize, we emphasize that the physical state (\ref{eq2a}) is found 
as a superposition of non physical solutions $\psi_a$. In the exact case,
these solutions would be degenerated, but they are split in a truncated Fock space
due to the effective violation of rotational invariance.  
Taking as a solution a superposition satisfying the 
angular condition -- in spite of the mass split, we restore the general   
properties of the wave function, violated by the truncation of the Fock 
space. This solution, which fulfills the 
correct transformation laws under rotation, approaches the exact one.
Our method thus provides a high accuracy solution of the relativistic 
two-body problem.\\   
\begin{figure}[htbp]  
\begin{center}  
\epsfxsize=7.5cm \epsfysize=9cm\subfigure[]{\epsffile{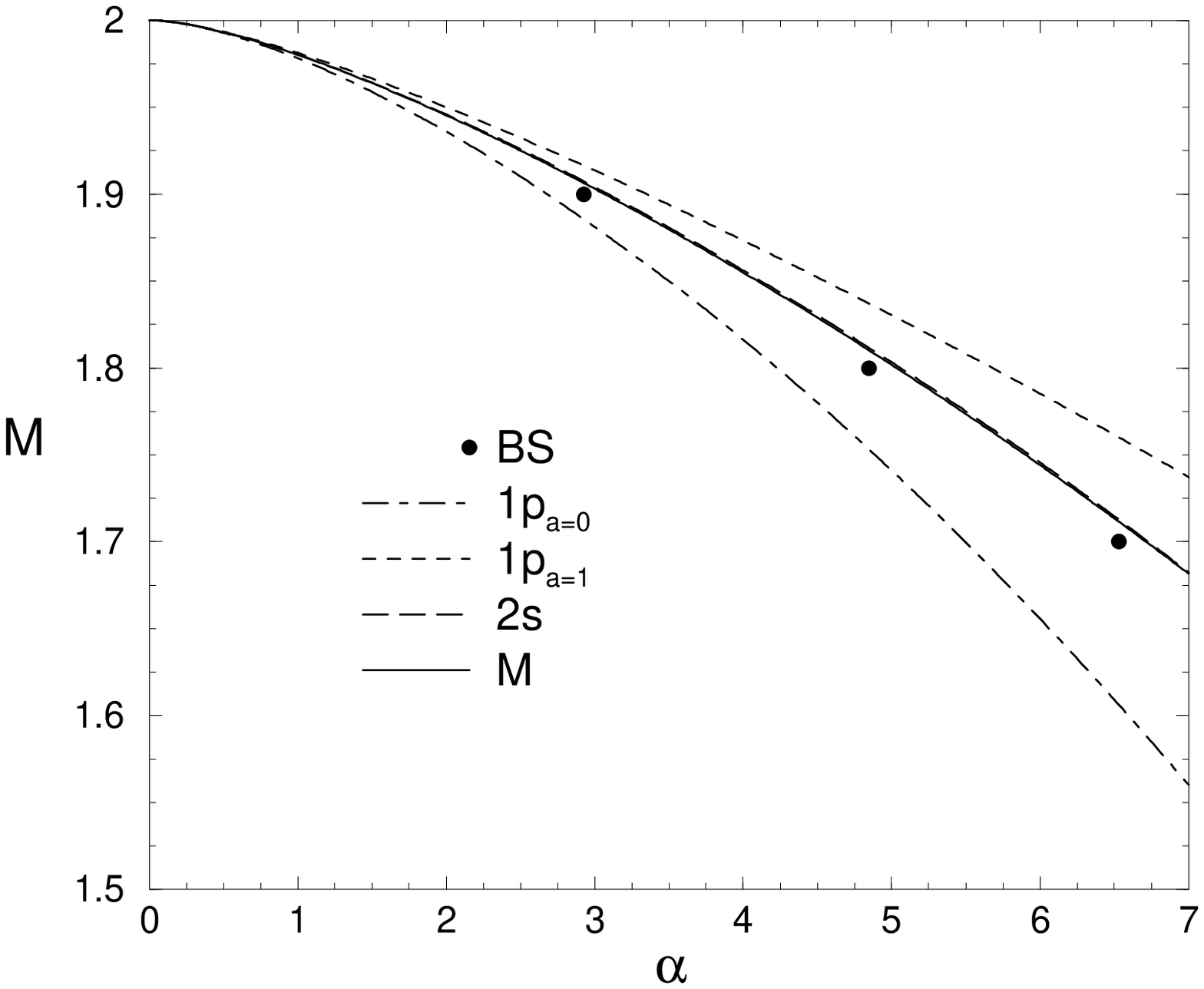}}
\label{2s1p_Mphys} \hspace{0.3cm}  
\epsfxsize=7.5cm \epsfysize=9cm\subfigure[]{\epsffile{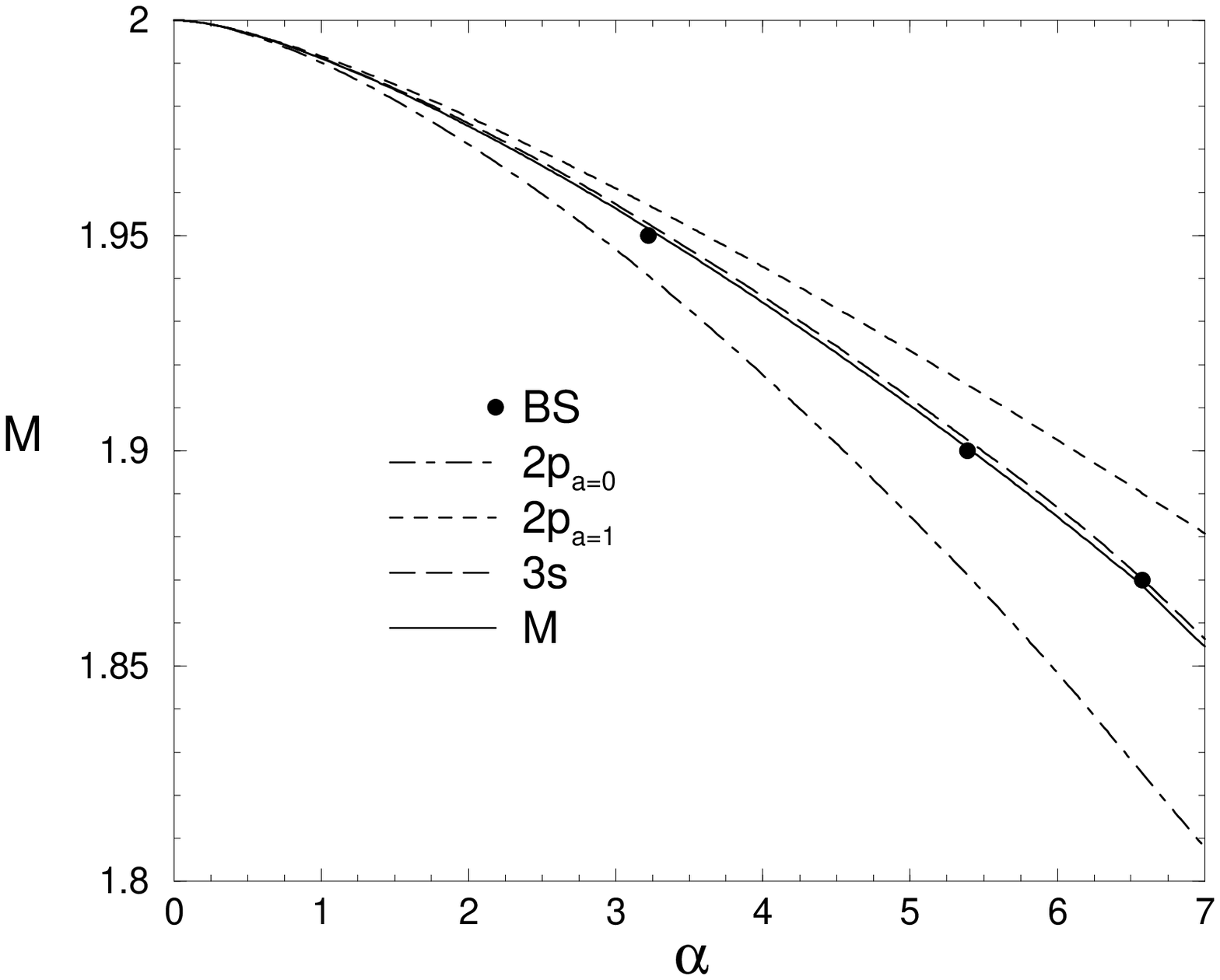}}
\label{3s2p_Mphys}  
\vspace{-1cm}
\end{center}  
\end{figure} 

%

\end{document}